\PassOptionsToPackage{svgnames,table,xcdraw}{xcolor}
\documentclass[sigconf,authorversion]{acmart}
\usepackage[nolist,nohyperlinks]{acronym}
\usepackage{multirow}
\usepackage[hang]{subfigure}

\setcopyright{rightsretained}

\copyrightyear{2023}
\acmConference[MMSys '23]{the 14th ACM Multimedia Systems Conference}{June 7--10, 2023}{Vancouver, BC, Canada}
\acmDOI{10.1145/3587819.3590971}
\acmISBN{979-8-4007-0148-1/23/06}

\begin{document}
\fancyhead[LE]{}
\fancyhead[RO]{}

\makeatletter
\renewcommand{\@copyrightowner}{BBC}
\makeatother

\title{Latency Target based Analysis of the DASH.js Player}

\author{Piers O'Hanlon}
\orcid{0009-0008-7462-1180}
\affiliation{
  \institution{BBC R\&D}
  \city{London}
  \country{United Kingdom}
}
\email{piers.ohanlon@bbc.co.uk}

\author{Adil Aslam}
\orcid{0009-0006-2915-5331}
\affiliation{
    \institution{BBC R\&D}
  \city{London}
  \country{United Kingdom}
}
\email{adil.aslam@bbc.co.uk}

\begin{abstract}
We analyse the low latency performance of the three \ac{abr} algorithms in the \texttt{dash.js} \ac{dash} player with respect to a range of latency targets and configuration options. We perform experiments on our \ac{dash} Testbed which allows for testing with a range of real world derived network profiles.  Our experiments enable a better understanding of how latency targets affect \ac{qoe}, and how well the different algorithms adhere to their targets. We find that with \texttt{dash.js} v4.5.0 the default Dynamic algorithm achieves the best overall \ac{qoe}. We show that whilst the other algorithms can achieve higher video quality at lower latencies, they do so only at the expense of increased stalling. We analyse the poor performance of L2A-LL in our tests and develop modifications which demonstrate significant improvements. We also highlight how some low latency configuration settings can be detrimental to performance.
\end{abstract}



\newpage

\maketitle

\begin{acronym}
  \acro{abr}[ABR]{Adaptive Bitrate}
  \acro{api}[API]{Application Programming Interface}
  \acro{bbc}[BBC]{British Broadcasting Corporation}
  \acro{cbr}[CBR]{Constant Bitrate}
  \acro{cdn}[CDN]{Content-Delivery Network}
  \acro{cmaf}[CMAF]{Common Media Application Format}
  \acro{cwnd}[CWND]{congestion window}
  \acro{dash}[DASH]{Dynamic Adaptive Streaming over HTTP}
  \acro{dvb}[DVB]{Digital Video Broadcasting}
  \acro{ewma}[EWMA]{exponentially-weighted moving average}
  \acro{ffi}[FFI]{Foreign Function Interface}
  \acro{has}[HAS]{HTTP Adaptive Streaming}
  \acro{http}[HTTP]{Hyper Text Transfer Protocol}
  \acro{ip}[IP]{Internet Protocol}
  \acro{ml}[ML]{machine learning}
  \acro{mpd}[MPD]{Media Presentation Description}
  \acro{mos}[MOS]{Mean Opinion Score}
  \acro{mss}[MSS]{Maximum Segment Size}
  \acro{mtu}[MTU]{Maximum Transmission Unit}
  \acro{os}[OS]{operating system}
  \acro{qoe}[QoE]{quality of experience}
  \acro{rtt}[RTT]{round-trip time}
  \acro{swma}[SWMA]{sliding-window moving average}
  \acro{sdn}[SDN]{Software Define Network}
  \acro{tbf}[TBF]{token bucket filter}
  \acro{tcp}[TCP]{Transmission Control Protocol}
  \acro{tv}[TV]{television}
  \acro{uk}[UK]{United Kingdom}
  \acro{url}[URL]{Uniform Resource Locator}
  \acro{vod}[VOD]{video on-demand}
  \acro{w3c}[W3C]{World Wide Web Consortium}
  \acro{whatwg}[WHATWG]{Web Hypertext Application Technology Working Group}
  \acro{xml}[XML]{Extensible Markup Language}
\end{acronym}

\section{Introduction}
\label{sec:intro}

For live content being distributed via both traditional broadcast and IP streaming, it would be desirable for IP services to at least match the latency of broadcast. There are a number of approaches to provide for such low latency IP streaming but their performance at different latency targets is not well studied. It is important to understand the appropriate compromise between delivery latency and \ac{qoe}.

In this paper we analyse the low latency performance of the widely used web-based \ac{dash} player, \texttt{dash.js}~\cite{dashjs}, examining a range of latency targets and configuration options, across its three \ac{abr} algorithms.  The default \ac{abr} algorithm, known as Dynamic\cite{spiteriTheoryPracticeImproving2018}, is a hybrid utilising its Throughput algorithm, which is based upon measured throughput, when the playback buffer is short, and utilising its Buffer Occupancy-based Lyapunov Algorithm (BOLA)\cite{spiteriBOLANearoptimalBitrate2016} at other times.

The Learn2Adapt Low Latency (L2A-LL)~\cite{karagkioulesOnlineLearningLowlatency2020} algorithm uses convex optimisation to predict the best option for future segment representations based on the impact the previous segment had on latency. The Low on Latency (LoL+)~\cite{limWhenTheyGo2020, bentalebCatchingMomentLoL2022} algorithm provides a \ac{qoe} maximisation decision making solution which utilises both a heuristic predictive model and a learning model based on self-organising maps. Each segment boundary is used as an opportunity to derive a predicted highest \ac{qoe} option out of the given representations, based on an internally defined weighted \ac{qoe} model.

Through a series of experiments on our DASH Testbed, we elucidate the trade-offs that may be made between Latency and \ac{qoe} when operating low latency streaming. Our Testbed allows for metrics collection with trace driven network emulation between a browser based player and a low latency streaming server, using chunked transfer mode with 3.84s segments. The metrics allow us to analyse the performance with respect to the factors such latency, video quality, and stalls. We employ the ITU-T P.1203 \ac{qoe} model and implementation \cite{Robitza2017d} to generate an overall \ac{qoe} score. We also examine the effect of other configuration options such as the FastSwitching algorithm \cite{spiteriTheoryPracticeImproving2018} which controls whether a player will try to obtain a replacement higher quality segment.

Whilst other analyses have focussed on achieving as low a latency as possible we find that at lower latencies ($\sim$3s) the \ac{qoe} is significantly lower than at higher latencies, but begins to level out around 8s. We make the following contributions:
\vspace{-\topsep}
\begin{itemize}
 \item A latency target based comparison of \texttt{dash.js}'s three low-latency algorithms.
 \item Find that with \texttt{dash.js} v4.5.0 default Dynamic algorithm achieves the best overall QoE and adherence to its latency targets.
 \item Show that the default enabled FastSwitching algorithm wastes bandwidth and does not generally improve the overall QoE.
 \item Elucidate the poor behaviour of the L2A-LL algorithm and demonstrate the benefits of our developed improvements.
 \item Demonstrate that Dash.js versions exhibit notable performance differences. 
\end{itemize}
\vspace{-\topsep}
Other work has mainly studied the performance of individual low latency ABR algorithms at fixed low latencies, tending to focus on the lowest possible latency without consideration for performance at higher latencies \cite{karagkioulesOnlineLearningLowlatency2020,limWhenTheyGo2020}. Also there are not many studies comparing \texttt{dash.js}'s three low latency \ac{abr} algorithms. One recent study \cite{zhangPerformanceLowLatencyDASH2022} performs a comparison but does not examine differing latency targets.

\begin{figure*}[ht!]
  \centering
  \includegraphics[width={0.65\textwidth}]{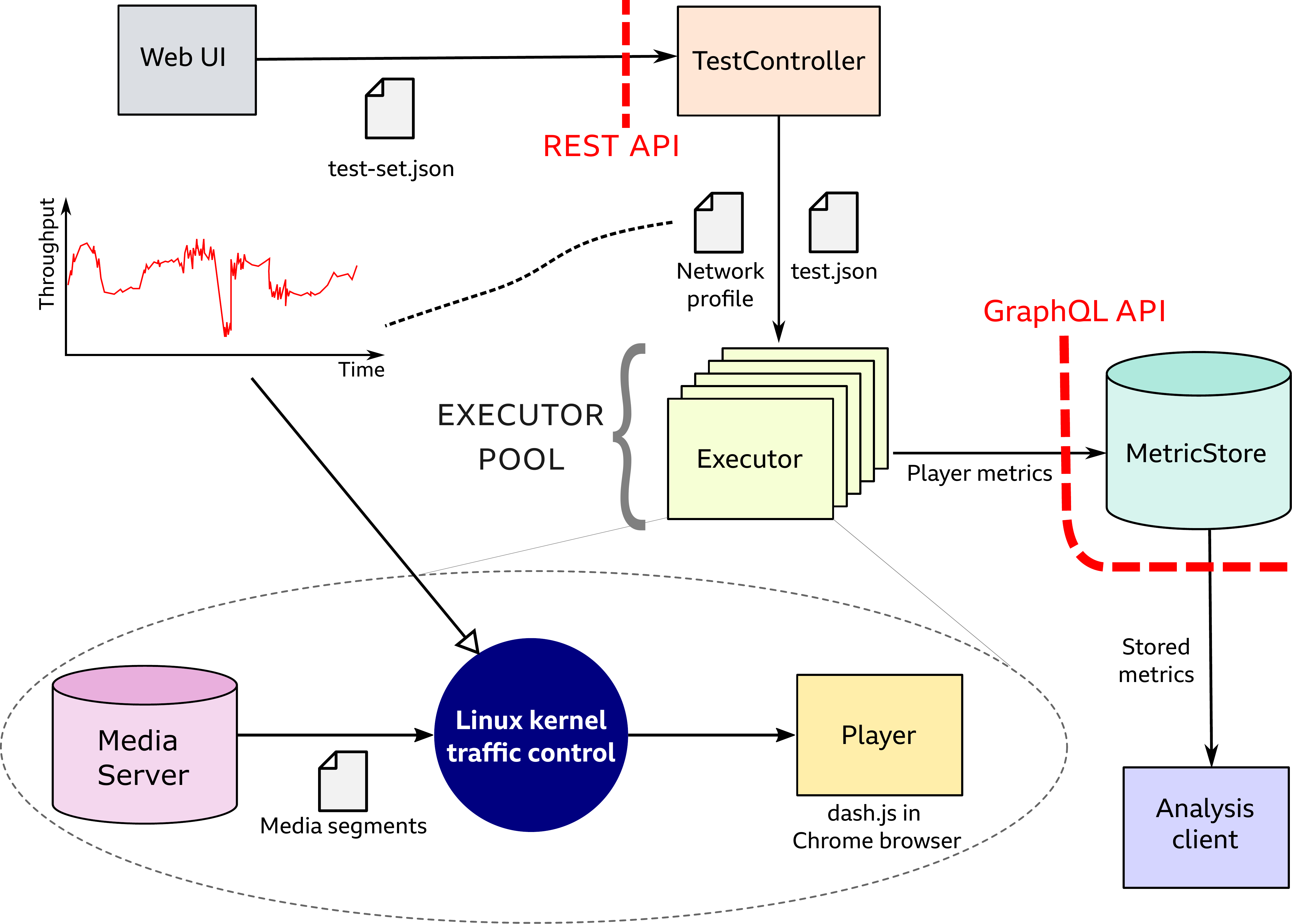}
    \caption{DASH Testbed architecture}
  \label{fig:exp-testbed} \end{figure*}

In many situations it is desirable to have a common latency across delivery mechanisms, platforms and devices, rather than the lowest latency achievable. This is important to viewers who might end up overhearing cheers for an event in a sports match. It also helps ensure that reaction on social media is similarly timed for everyone. Since latency, reliability and quality are interrelated, seeking minimum latency inevitably reduces reliability or quality or both so it is important to consider what latency is actually needed for a given situation and not go further than is necessary.

The remainder of this paper is structured as follows: first, in §\ref{sec:experiments} we present details of our experimental testbed and methodology, next, in §\ref{sec:results} we present our results, and in §\ref{sec:further-analysis} we provide further analysis and recommendations. We summarise related work in §\ref{sec:related}, and finally, we conclude with our key findings and provide potential avenues for future research in §\ref{sec:conclusions}.

\section{Experiments}
\label{sec:experiments} We evaluated the performance of \texttt{dash.js} using our DASH Testbed \cite{DASHTestbed}. We chose to develop our own testbed, as opposed to using others, so we had full control over all aspects of its behaviour.  We ran a series of experiments with four different network profiles, emulating a range of network conditions, enabling us to compare the performance of the three ABR algorithms.

\subsection{DASH Testbed}
\label{ssec:exp-testbed} 

\begin{table*}[!ht]
  \centering
  \begin{tabular}{lccccccccc}
\toprule
Resolution     & 192x108 & 256x144 & 384x216 & 512x288 & 704x396 & 896x504 & 704x396 & 960x540 & 1280x720 \\ Bitrate (Kbps) & 86      & 156     & 281     & 437     & 827     & 1,374   & 1,570   & 2,811   & 5,468 \\
    Frame Rate     & 25      & 25      & 25      & 25      & 50      & 25      & 50      & 50      & 50 \\
    \bottomrule
  \end{tabular}
  \caption{The video bitrate ladder used for our experiments}
  \label{tab:exp-representations} \end{table*}

\begin{figure*}[!ht]
  \centering
    \includegraphics[width={0.95\textwidth}]{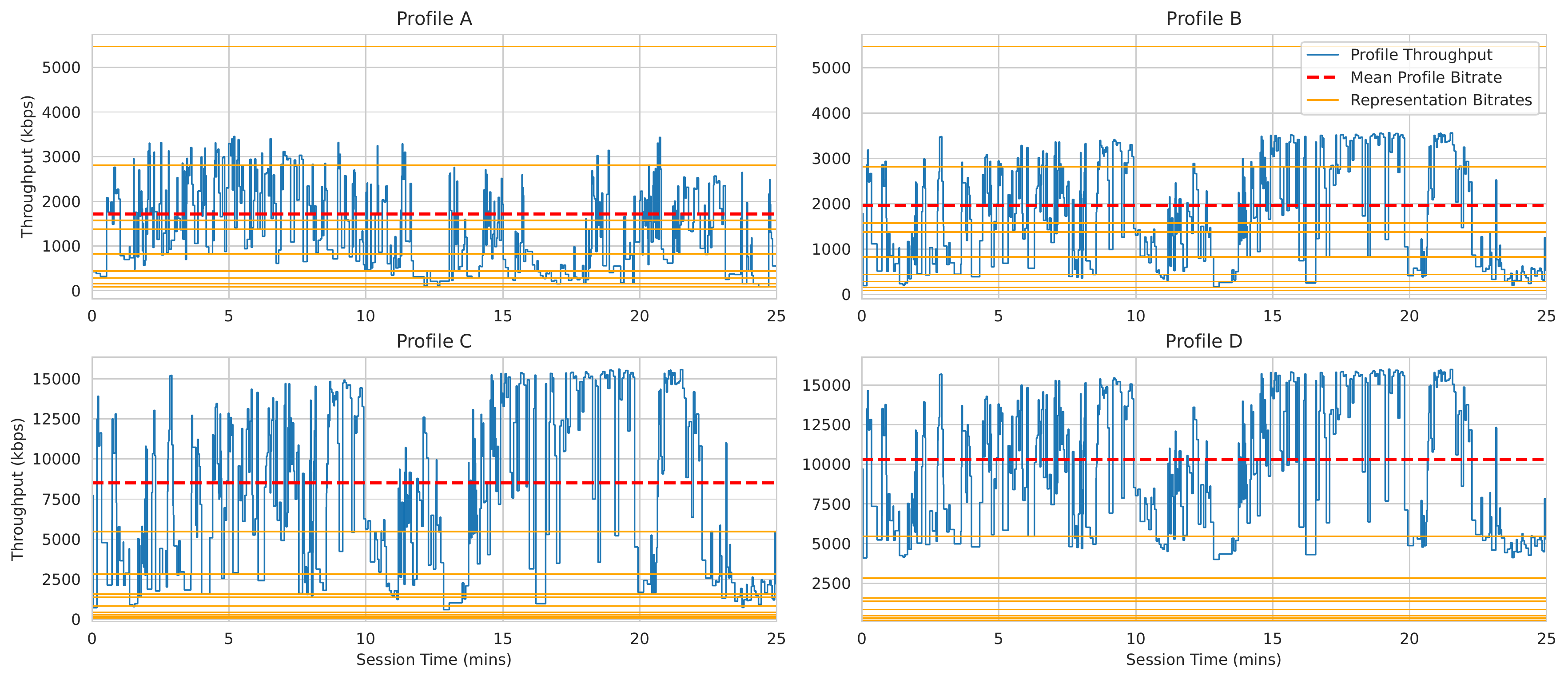}
    \caption{Network Profiles (with bitrate ladder)}
  \label{fig:exp-Networkprofiles} \end{figure*}

Figure~\ref{fig:exp-testbed} displays the overall architecture of the DASH Testbed.  At the heart of the Testbed sits the TestController, which is responsible for queueing and running sets of tests. A web based UI allows for submission of test sets to the TestController in a JSON file, which describes each set of tests to be run. This JSON file allows for control over the configuration of the Testbed. Firstly, it allows specification of the stream to use in the test: whether it’s a live or on-demand stream, what combination of audio/video/subtitle components it has and its duration. Secondly, it configures which DASH player software to use in the test and how that player should be configured. The Testbed currently supports players based around the dash.js JavaScript DASH implementation and the GStreamer media toolkit. In our case we were utilising the Google Chrome browser running \texttt{dash.js}. Thirdly, one can set the number of sessions of that test to run. And finally it allows specification of the network profile that should be applied during the running of each test session.

When the TestController runs a test from within a test set, it farms out the individual sessions of that test to a pool of Executors, which are each responsible for running a single session at a time. Each of the Executors is a separate physical Linux machine so all the tests are fully independent of one another. We use the traffic control facilities built into the Linux kernel to throttle the data rate of its network interface according to the particular network profile selected for that test and we ensured that the interfaces utilise the usual \ac{mtu} of 1500 bytes. The Executor configures and launches the player identified in the test description, which will play the test stream in real time and report metrics about its performance back to the MetricStore. 

Metrics are obtained from both the in-built dash.js DVB DASH metric reporting capabilities and from custom JavaScript functions. Generally metric data types can be categorised into two groups: event based, and periodic. Event based metrics are reported at the time they occur typically with additional context, such as: HTTP transactions, rebuffering periods, and representation switches. The number of event-based metrics per run can therefore vary, and in some cases be zero. Periodic reporting is used to monitor certain metrics, such as buffer levels and latencies. Periodic reporting provides values of a parameter at regular intervals, typically 500 milliseconds, as set by the Testbed. These metrics typically have the same number of entries per run. All metrics are time-stamped.

The available video representations within the manifest can be seen in Table~\ref{tab:exp-representations}, which are representative of those used in the industry. The segments use \ac{cmaf} and have a length of 3.84 seconds, consisting of four 0.96 second chunks. The content consists of a subsection of news footage which has been encoded as H.264 \ac{cbr} streams and repackaged for use in the Testbed media server as live low latency encoded media. The test stream also includes a single 128kbps audio representation. The \ac{qoe} measure we utilise is not content dependent so the content genre used has no bearing on the results.

The network profiles in Figure~\ref{fig:exp-Networkprofiles} used in the test runs were derived from data captured from BBC iPlayer streaming sessions. The orange lines indicate the bitrates of the tested representations, illustrating how these compare to the throughput in the various network profiles.

\subsection{Methodology}
\label{ssec:exp-method} 

As the primary comparison for live IP streaming, typical terrestrial broadcast (DVB-T) latencies were used as a basis for selecting latency targets to be tested. This comparison is important for mass adoption, as users/consumers will expect a comparable product to the existing medium. 

A variety of latency targets were selected to overlap the existing latency range on DVB-T, which is typically between 6 - 8 seconds. The values chosen were 3, 5.5, 8, and 15 seconds. These first three values are equally spaced apart to provide measurement points across the range. The target of 15 seconds was also chosen to be used as a relatively stable operation point, which would also allow for a comparison with traditional (non-low-latency) live streaming.

The version of \texttt{dash.js} used was v4.5.0, released in September 2022. The configuration parameters for dash.js were mostly kept to default values. Certain parameters were changed to effect particular functionality, utilise existing playback choices, or to vary the test scenarios. We set the \emph{maxDrift} to 5s and \emph{playbackRate} to 0.17. The playback rate was set such that at maximum rate, the stream would catch up approximately 5 seconds within a 30 second window.

The catchup mechanism employed was set to the default option for the test cases utilising Dynamic, and L2A-LL ABR algorithms. The LoL+ algorithm utilises its own custom catchup mechanism so that was selected for its test cases only. 

For the throughput estimation technique the default "moof" parsing method was used, as opposed to the older data-chunks method. Sliding window was selected as the moving average technique, which is also the default method.

\begin{figure*}[!ht]
  \centering
    \includegraphics[width={1.0\textwidth}]{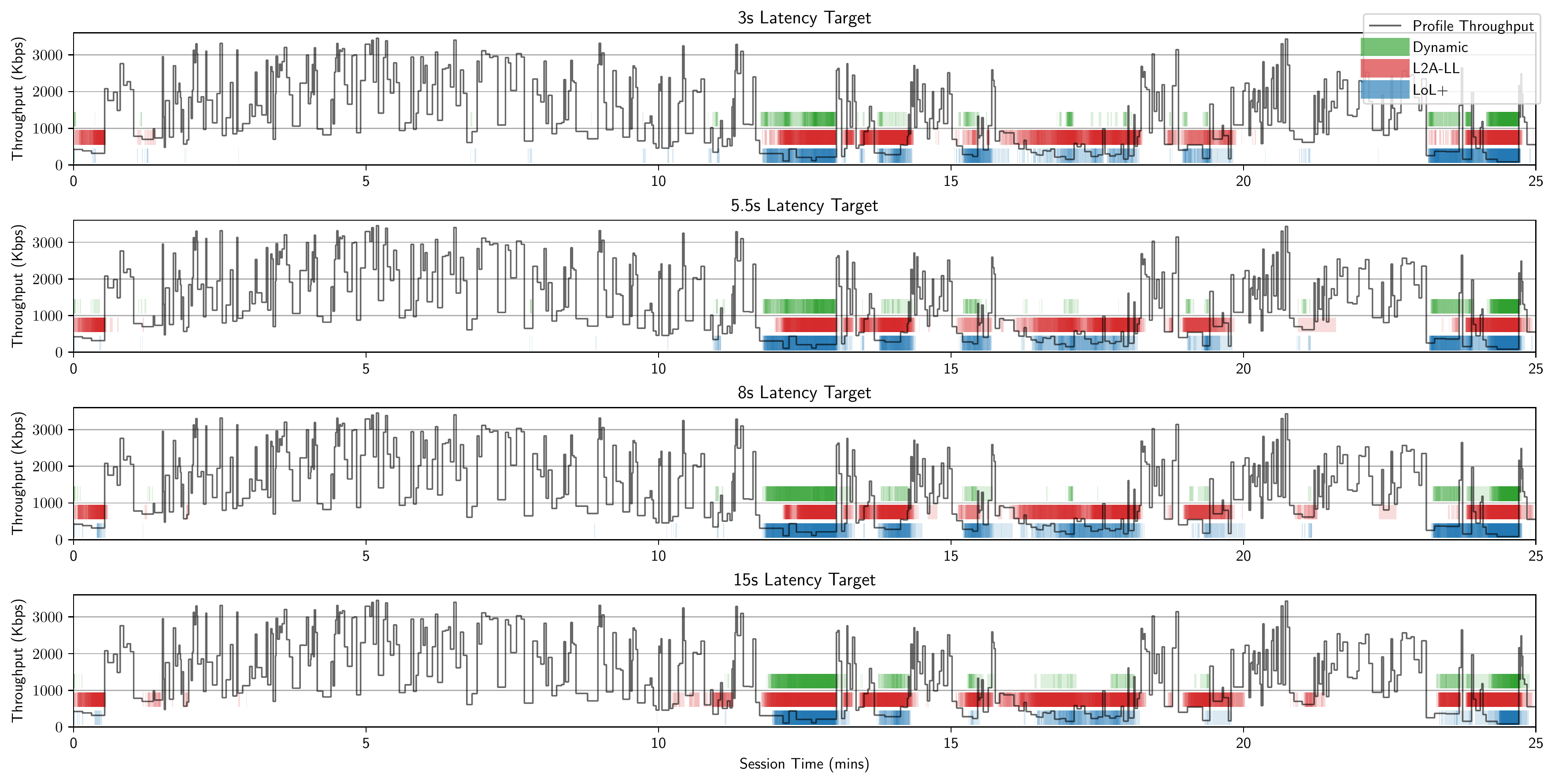}
    \caption{Stall Heatmap (Profile A)}
  \label{fig:results-stall_heatmap} \end{figure*}

To ensure each run would provide results independent of previous tests we set both \emph{lastBitrateCachingInfo} and \emph{lastMediaSettingsCachingInfo} to false. In normal use enabling these parameters can improve the start-up performance of a session as they allow for the use of previously acquired throughput estimation and representation selections as factors for deciding initial representation requests.

We ran all tests apart from the configuration comparison test with the FastSwitching feature disabled as it was found to adversely affect the performance (see Section~\ref{ssec:config-opts}).

A test scenario is defined by a specific latency target, with one of the three ABR algorithms, operating on one of the four network profiles. The Testbed was configured with an \ac{rtt} of 50ms.  Each test scenario would run for 1497.6 seconds (24 minutes and 57.6 seconds).  This particular duration comes from the number of segments that would be used, 390, assuming continuous playback at 1x playback rate of 3.84s duration segments.

Each test scenario was run twenty times. This was to ensure any averages calculated would have a sufficient level of confidence.

\section{Results}
\label{sec:results} 

In this section we provide an analysis of our core experiments using \texttt{dash.js} v4.5.0. First, we break it down according to the key metrics we examined, then consider the overall QoE.

\subsection{Rebuffering}
\label{ssec:results-rebuffering} 

The negative impact of rebuffering on \ac{qoe} is widely acknowledged for \ac{vod}\cite{dobrianUnderstandingImpactVideo2011, krishnanVideoStreamQuality2012, dimopoulosMeasuringVideoQoE2016}. In a low-latency context the negative impact of time spent rebuffering on \ac{qoe} is greater than in a \ac{vod} setting\cite{dobrianUnderstandingImpactVideo2011} as it increases the playback latency.

The amount of stalling each ABR algorithm experiences varies considerably between the three, in both number of stalls, and the duration of rebuffering that follows.

Figure~\ref{fig:results-stall_heatmap}, which shows a heatmap of the stalling, utilises a set of coloured bars indicating the stalls for each ABR algorithm overlaid onto the network profile. The Dynamic algorithm performs the best of three in this regard, with the least number of stalls and shortest time overall spent rebuffering. LoL+ follows with an increase in both metrics, and L2A-LL performs the worst  of the three, by a considerable margin.

Increasing the target latency does lead to reduced stalling levels as the larger buffer provides a greater period of time for the algorithms to adapt. These are reflected in their improving \ac{qoe} scores which may be found in Section~\ref{ssec:results-qoe}.

Whilst we do not have the space to show the other scenarios we see a similar trends across the different network profiles.

We expect that less time rebuffering enables clients to stay closer to the live-latency target, a measure that is evaluated next.

\begin{figure*}[!ht]
  \centering
    \subfigure[Profile A]{
      \includegraphics[width={0.48\textwidth}]{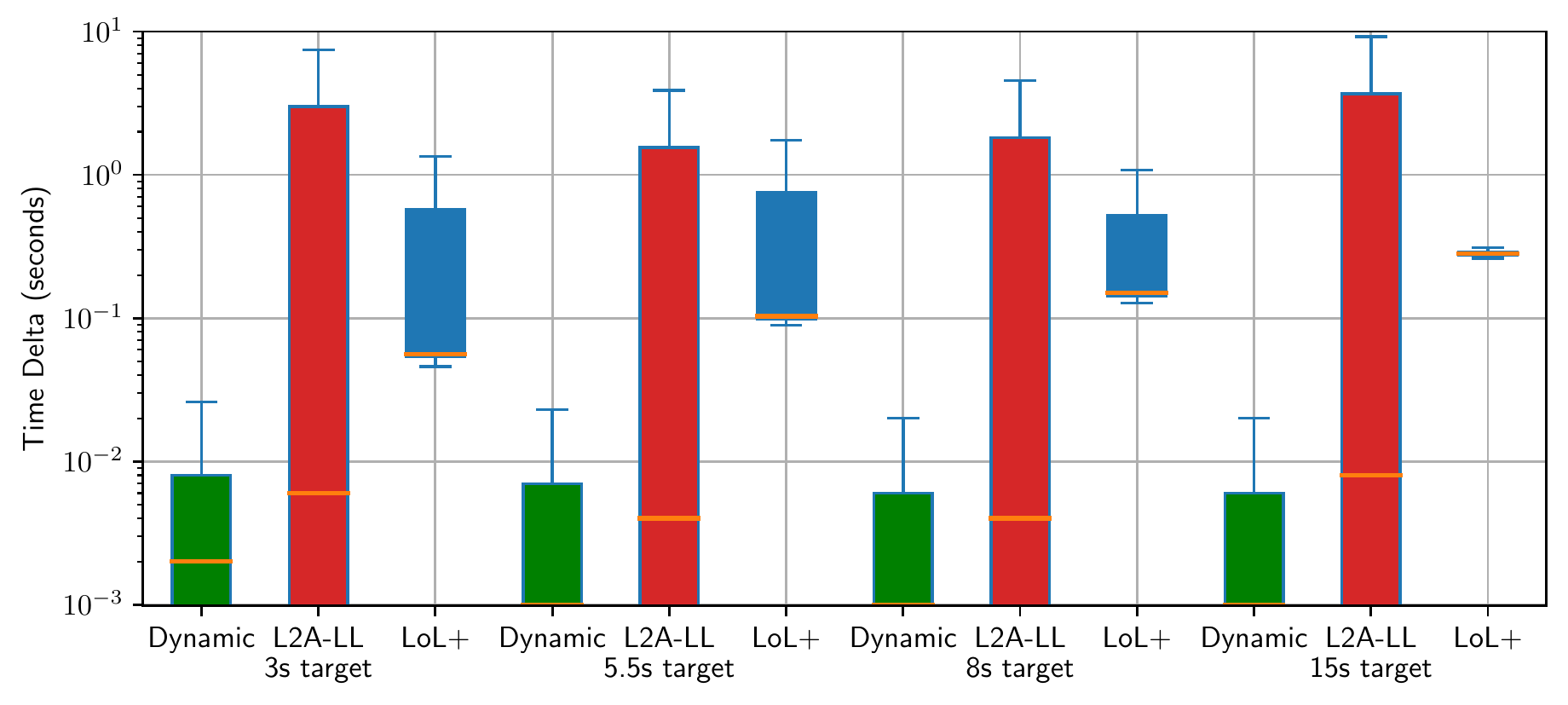}
    \label{fig:results-mean-dlatency-a} }
    \subfigure[Profile D]{
      \includegraphics[width={0.48\textwidth}]{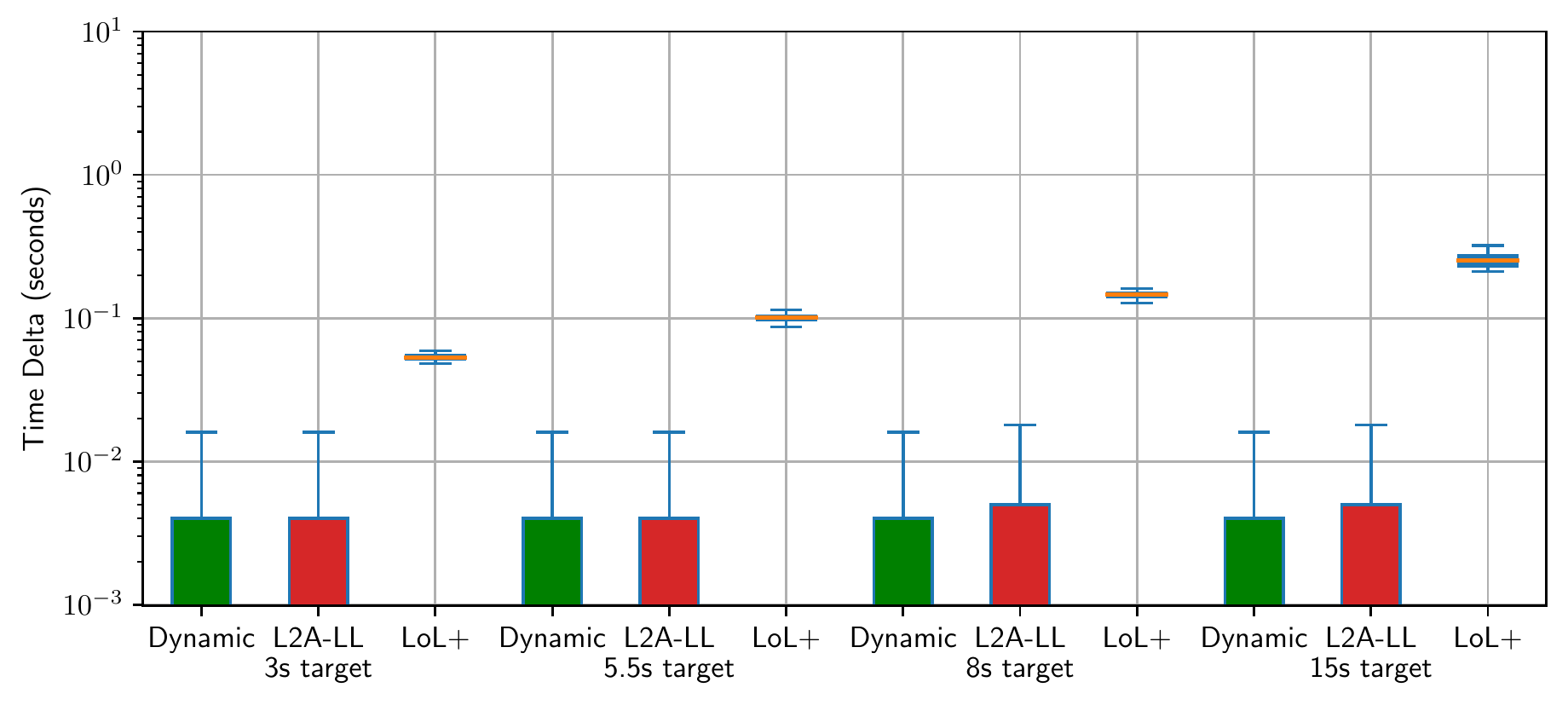}
    \label{fig:results-mean-dlatency-d} }
    \caption{Latency Deviation from Target}
  \label{fig:results-mean-dlatency} \end{figure*}

\begin{figure*}[!ht]
  \centering
    \subfigure[Profile A]{
        \includegraphics[width={0.47\textwidth}]{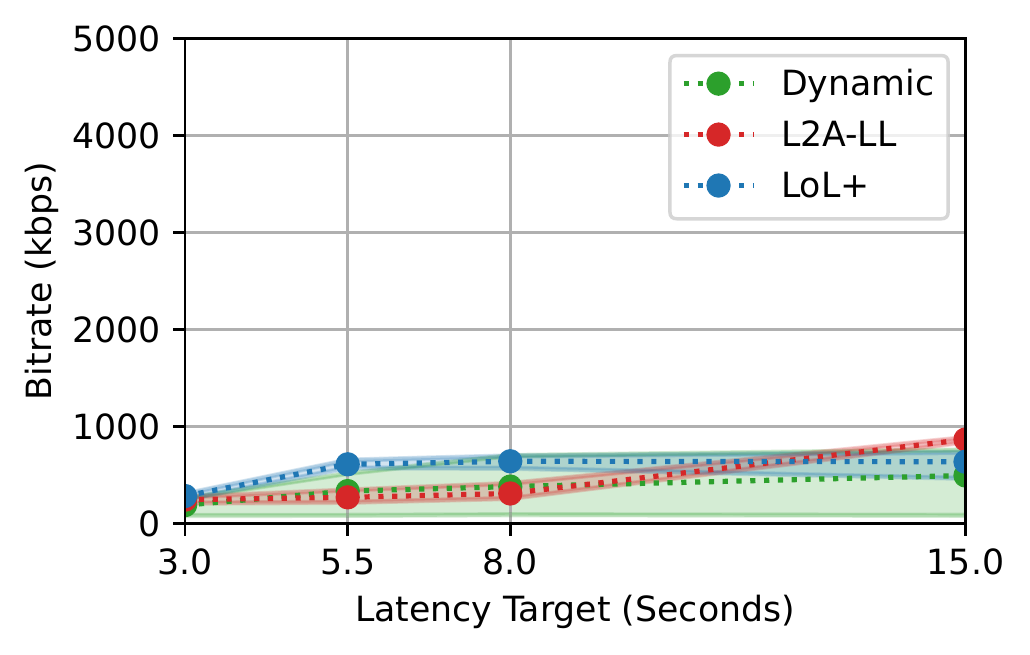}
    \label{fig:results-meanbitrate-a} }
    \subfigure[Profile D]{
        \includegraphics[width={0.47\textwidth}]{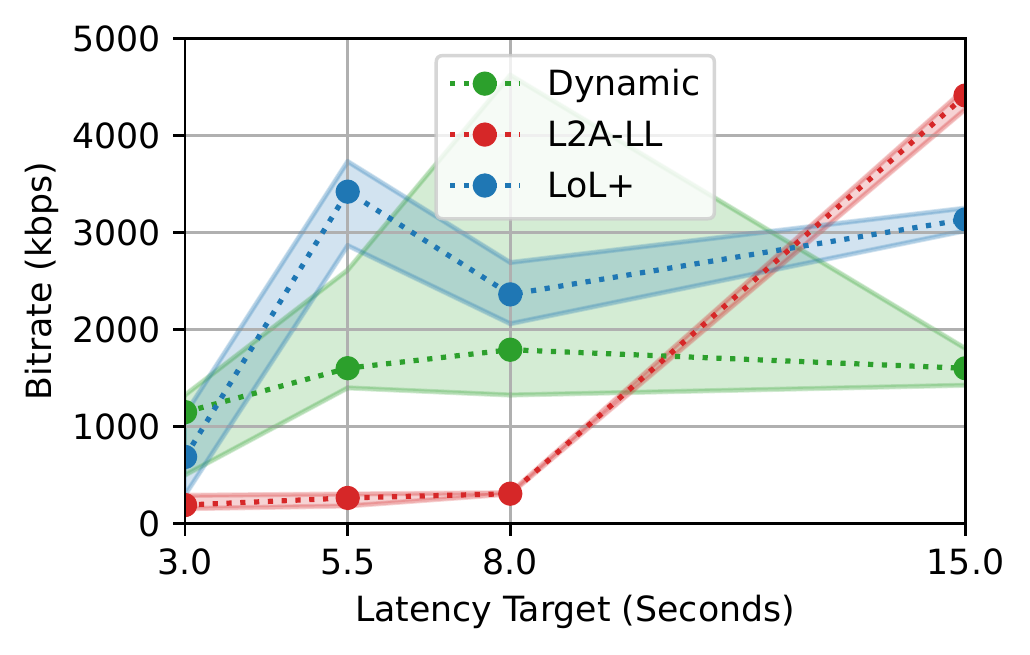}
    \label{fig:results-meanbitrate-d} }
    \caption{Mean Video Bitrates}
  \label{fig:results-meanbitrates} \end{figure*}

\subsection{Latency}
\label{ssec:results-latency} 

In this section we explore the performance at the range of tested latency targets. We explore how well the different algorithms adhere to their targets.

The measure of latency used here is the delay between when content has been made available at the origin to when it is rendered at the client. In low-latency streaming, a live-latency target is selected by the content provider and usually configured in a \ac{mpd}. Clients should stay at or close to the target so that the experience of viewing events on the screen is not pre-empted by social media, or neighbours reacting to the same event moments earlier. The live latency values are retrieved from the \texttt{dash.js} client at 500ms intervals.

In Figure~\ref{fig:results-mean-dlatency} we can see how the different algorithm adhere to their latency targets, using boxplots, where the median is marked by an orange line, with the whiskers extending to 1.5 x interquartile range. We use a log scale; where no median is shown it is zero in Figure~\ref{fig:results-mean-dlatency-d}. We observe that Dynamic provides the smallest deviation from the target in both scenarios. Figure~\ref{fig:results-mean-dlatency-a} shows the performance on the most challenging scenario, and the trend remains largely similar up to the least challenging profile, in Figure~\ref{fig:results-mean-dlatency-d}, though the confidence intervals narrow and the time delta decreases. We notice that there seems to be a small systematic increase in the delta for LoL+.

\subsection{Video Quality}
\label{ssec:results-quality} 

The selected video bitrate is regarded as quite influential at predicting user satisfaction for video-on-demand~\cite{dobrianUnderstandingImpactVideo2011}. We first evaluate this measure, and follow with further results that suggest bitrate, alone, may be misleading or incomplete measure in the context of live-streaming.

As we can see from Figure \ref{fig:results-meanbitrates}, where we plot the mean across the 20 runs and the shaded area is defined by the 5th and 95th percentiles, they vary significantly between the algorithms. In the most challenging scenario, in Figure \ref{fig:results-meanbitrate-a}, they all struggle to maintain a rate higher than 1000Kbps, whilst with the least challenging situation, in Figure \ref{fig:results-meanbitrate-d}, the bitrates of the three algorithms generally fluctuate significantly between each latency target, with Dynamic providing the most stable but lower bitrate, whilst L2A-LL and LoL+ reach a higher bitrate.

We calculate the mean video bitrate ($\overline{B}$) using Equation \ref{eq:results-bitrate}. We remove any duplicate requests from the representation download count.

\begin{align}
  \label{eq:results-bitrate}
    & \overline{B} = \frac{segment\_{duration}}{total\_{playback}\_{time}} \sum_{R=108p25}^{720p50} (N_{R} \times B_{R}) \\
    & Where\ R = Representation \nonumber \\
    & N_{R} = Number\ of\ times\ R\ downloaded  \nonumber \\
    & B_{R} = Bitrate\ of\ Representation \nonumber
\end{align}

\begin{figure*}[!ht]
  \centering
    \subfigure[Profile A]{
        \includegraphics[width={0.32\textwidth}]{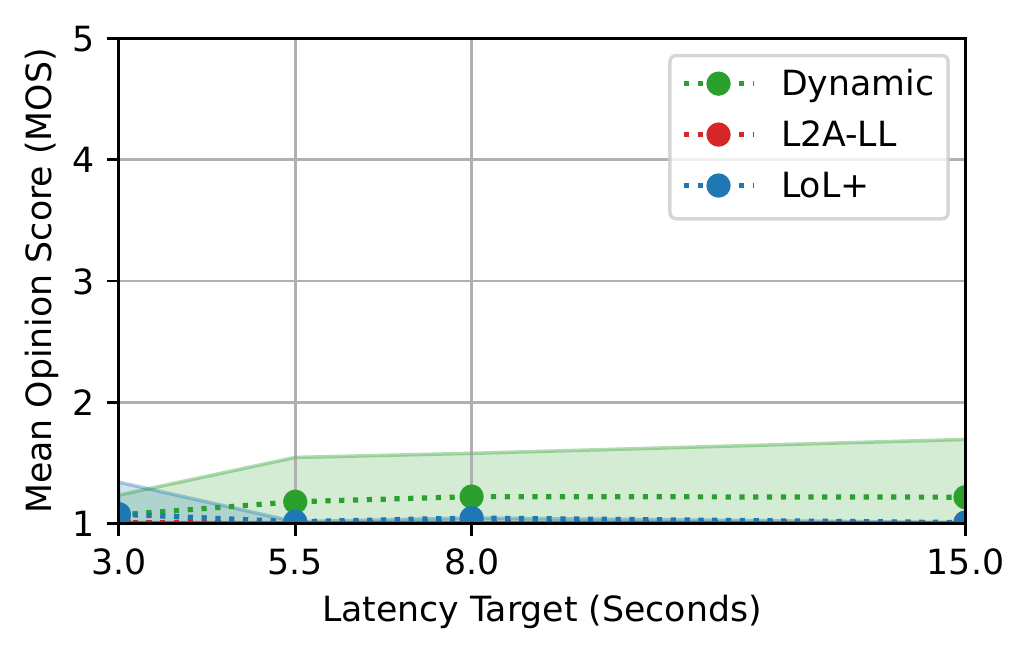}
        \label{fig:results-qoe-a} }
    \subfigure[Profile C]{
        \includegraphics[width={0.32\textwidth}]{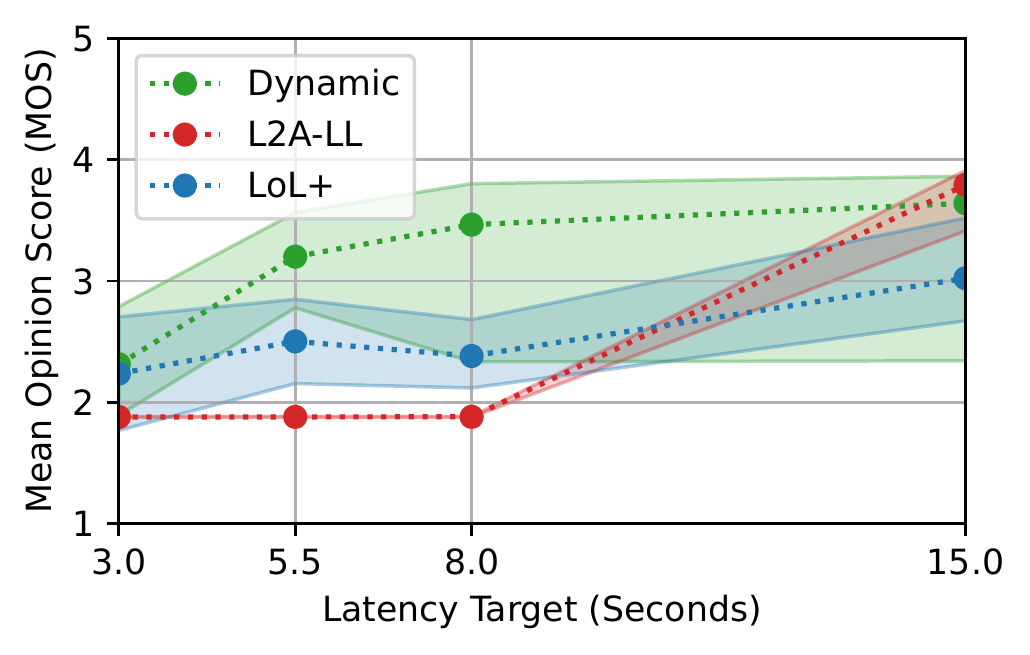}
        \label{fig:results-qoe-c} }
    \subfigure[Profile D]{
        \includegraphics[width={0.32\textwidth}]{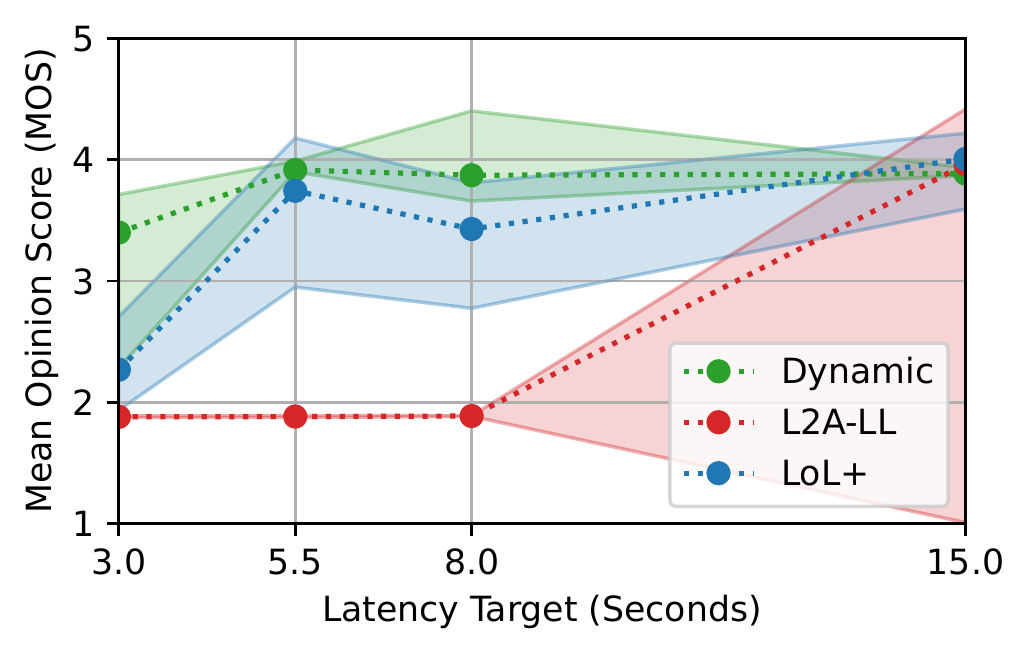}
        \label{fig:results-qoe-d} }
    \caption{QoE - P.1203 overall measure}
    \label{fig:results-qoe-p1203} \end{figure*}

\begin{figure*}[!ht]
  \centering
    \subfigure[Modified L2A-LL]{
        \includegraphics[width={0.32\textwidth}]{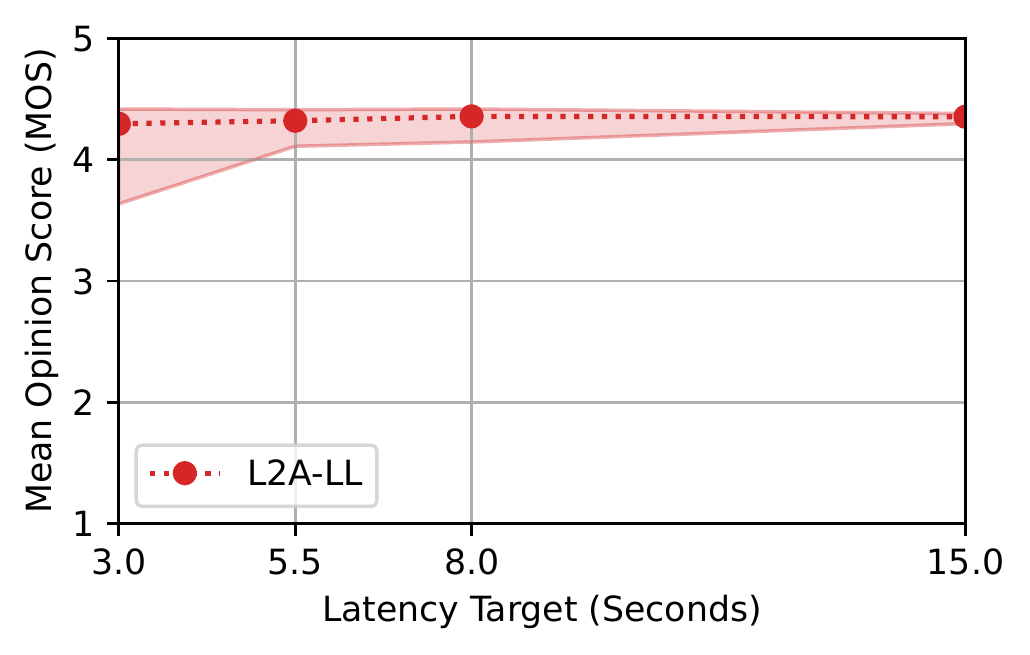}
        \label{fig:results-qoe-L2A-SKIPinit} }
    \subfigure[FastSwitching Enabled]{
        \includegraphics[width={0.32\textwidth}]{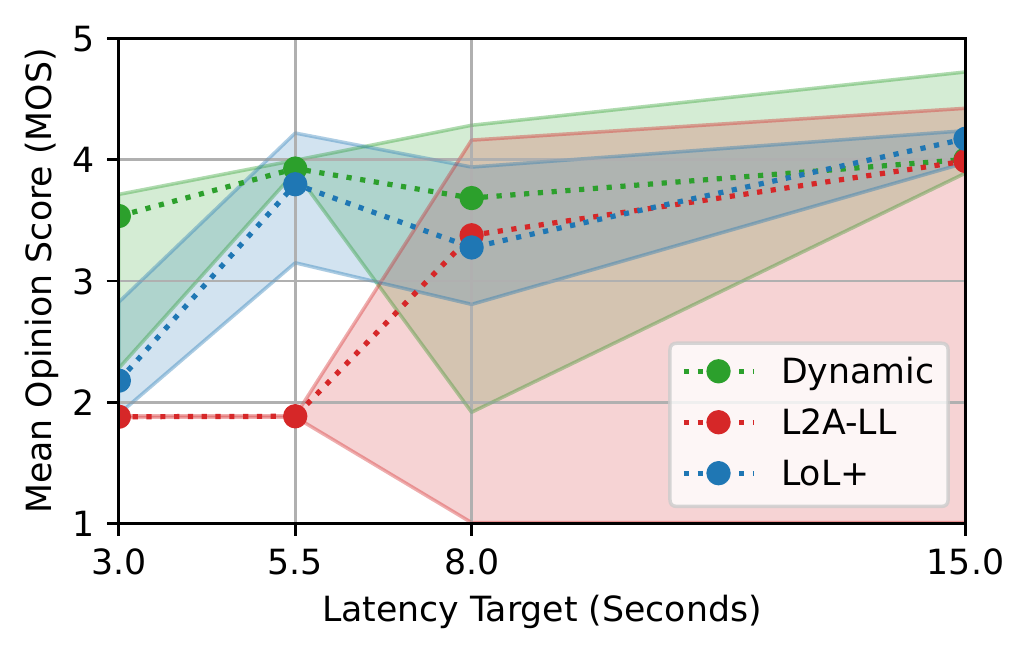}
        \label{fig:results-qoe-fs-on} }
    \subfigure[dash.js v4.3.0]{
        \includegraphics[width={0.32\textwidth}]{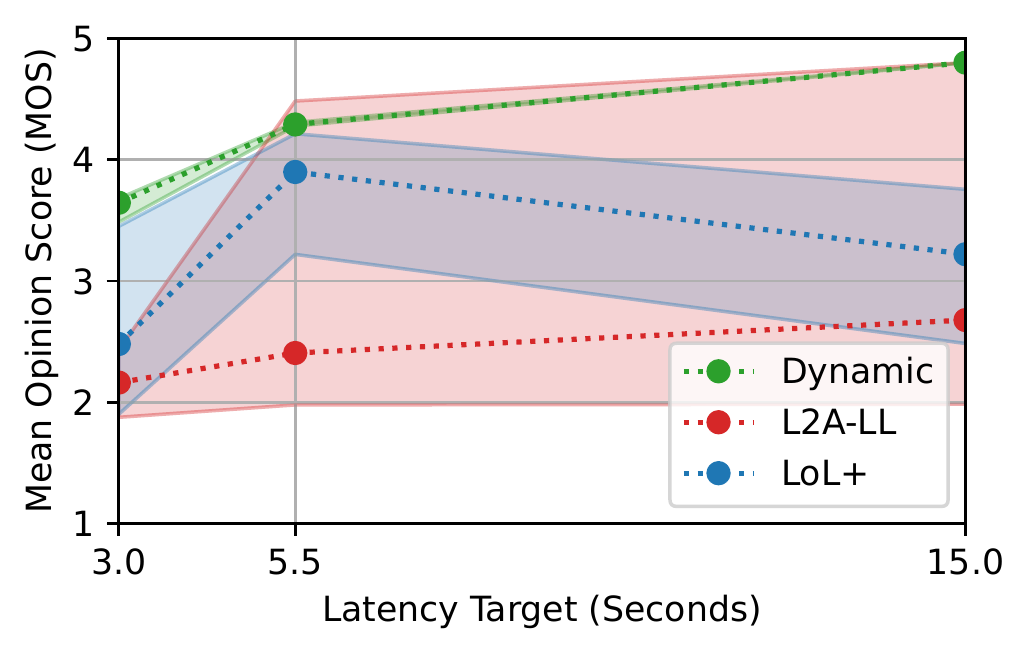}
        \label{fig:results-qoe-d-430} }
    \caption{QoE - P.1203 overall measure: Further analysis (Profile D)}
    \label{fig:results-qoe-p1203-comp} \end{figure*}

\subsection{Quality of Experience}
\label{ssec:results-qoe} 

We utilised ITU-T Rec. P.1203, a more comprehensive quantitative measure of overall \ac{qoe}, which provides for parametric and machine learning bitstream-based quality assessment of adaptive audiovisual streaming services over reliable transports. We used the reference implementation~\cite{Robitza2017d} to generate an overall \ac{qoe} score, in the form of a \ac{mos}, using the Mode 0 which is based upon stream metadata. We employ this \ac{qoe} measure as it has been extensively trained and validated with subjective tests on over a thousand audiovisual sequences containing HAS-typical effects (such as stalling, representation switches) and it is independent of absolute latency, though it does respond to changes in latency through its stalling measure. In contrast the low latency specific \ac{qoe} measures used in other work~\cite{yiACMMultimedia20192019,bentalebCatchingMomentLoL2022} lack a basis in subjective user testing and also contain a latency factor which rewards them according their lower absolute latency, as opposed to how well an algorithm adheres to its latency target.

In Figure~\ref{fig:results-qoe-p1203} we plot the P.1023 QoE overall results with the points indicating the mean across the 20 runs and the shaded area is defined by the 5th and 95th percentiles. Whilst the results are all quite low with Profile A, in Figure~\ref{fig:results-qoe-a}, we see that Dynamic is on top, but with Profiles C, Figure~\ref{fig:results-qoe-c}, and D, Figure~\ref{fig:results-qoe-d}, the Dynamic algorithm out performs the other algorithms at all latency targets until they all converge at the top 15s target.

\section{Further analysis}
\label{sec:further-analysis}
In this section, we explore some features of our results and discuss why they arise, proposing improvements to \texttt{dash.js} where appropriate.

\begin{table*}
\resizebox{\textwidth}{!}{\begin{tabular}{lllllllllllll}
    \toprule
    \ & 3s Latency & Target  &&  5.5s Latency & Target   & & 8s Latency & Target   & & 15s Latency  & Target   \\
    \ & Dynamic &  L2A-LL & LoL+ & Dynamic &  L2A-LL & LoL+ & Dynamic &  L2A-LL & LoL+  & Dynamic &  L2A-LL &  LoL+ \\
    \midrule
    Mean unique segments          & 391.4 & 124.5 &391.2& 392.0   & 392.0 & 392.0 & 392.6 &   385.6 &  392.4 &  393.9 &  393.6 &    393.6 \\
    Mean total segments           & 391.4 & 126.6 &391.2& 392.0   & 392.0 & 392.0 & 397.3 &  484.7 &  436.0 &  394.7 &  425.8 &      424.0 \\
\textbf{Rerequest percentage} & 0.0\% &  1.63\% &  0.0\%  & 0.0\% &  0.0\% &  0.0\%   & 1.1\% &  20.4\% &   10.0\% &        0.2\% &       7.5\% &    7.1\% \\
    \bottomrule
\end{tabular}}
    \caption{Segment requests with FastSwitching enabled on Profile D}
  \label{tab:exp-rerequests} \end{table*}

\subsection{L2A-LL analysis and improvements}

\label{ssec:results-qoe-L2A} 

We took a closer look at why L2A-LL was performing so poorly below the 15s target (in Figure~\ref{fig:results-qoe-d}) and discovered that a combination of factors were coming to into play. Since we are testing with chunk durations of 960ms \texttt{dash.js} spends more time waiting for the segments to become available than with shorter chunks, and in low latency mode the scheduler has a timeout of 300ms so the ABR algorithms can be called multiple times before the new segment becomes available. This is doesn't appear to be a problem for the other algorithms but with L2A-LL it can lead to the algorithm's state being unintentionally updated many times in quick succession for each segment. This is compounded by an issue specific to L2A-LL where it utilises the most recent segment's throughput estimate as an input to the algorithm but fails to exclude measurements derived from initialisation segments. The problem with initialisation segments is that the throughput estimates are generally very low due to their small size ($\sim$900 bytes) and difficulty in accurately measuring their delivery time - indeed they are excluded from \texttt{dash.js}'s throughput history. For the streams below the 15s target it is these problems that act together which leads L2A-LL to get stuck in a low state: When it first tries to switch up to a higher bitrate Representation, this first causes an initialisation segment to be downloaded.  When dash.js then tries to obtain the next media segment, it is yet to become available and the problems described mean that the L2A-LL algorithm is run again, this time using the initialisation segment’s low throughput estimate.  This drives the algorithm to switch down again, giving poor a quality of experience. The situation is different in the case of the 15s target as it exceeds the default 12s value of the configuration parameter \emph{stableBufferTime} which limits the size of the buffer that \texttt{dash.js} builds when it is not at the top quality.  In this case it is requesting segments behind the leading edge when they are fully available and any throughput estimates from initialisation segments are replaced with representative measurements from media segments before the L2A-LL algorithm runs.

We developed modifications to the L2A-LL algorithm to address the problems we highlighted by limiting the algorithm to only update its state if the throughput estimate comes from a segment that it hasn’t used before and is not an initialisation segment.  In other cases, it returns its previously calculated representation quality switch decision. As can be seen in Figure~\ref{fig:results-qoe-L2A-SKIPinit} the modified L2A-LL implementation performs significantly better at all latency targets. These modifications will be contributed back to the \texttt{dash.js} project.

\subsection{Configuration Options}
\label{ssec:config-opts} 

We studied the effect of other configuration options including the FastSwitching algorithm \cite{spiteriTheoryPracticeImproving2018} which controls whether a player will try to obtain a replacement higher quality segment.

To understand the behaviour of this feature we examined the requests for segments, comparing the ratio of unique segments with the total expected number of segments (390, as explained in Section~\ref{ssec:exp-method}) to the percentage of time during a session which playback wasn’t stalled.  Since these are tests with a live stream, within a specified time limit (1497.2s), if they undergo sufficient stalling not all the segments will be requested, and some cases there can be slightly more segments requested due to good performance and playback speed increases.  We utilise the HTTP transactions list, from the DASH metric reports, to identify the number of segments for a run and the number of unique segments downloaded. 

When the FastSwitching feature is enabled, which it has been by default since \texttt{dash.js} v4.0, it was expected that the algorithms would look to switch up to a higher bitrate representation, if the buffer is larger than the threshold and there is sufficient network throughput, replacing the previously downloaded segment before it is played. This was expected to only affect a small number of segments during a session, and mostly in higher latency target scenarios, where a large buffer increases the download-to-playout time.

With FastSwitching enabled the buffer threshold level is set to 1.5 $\times$ segment\_{duration}\footnote{The segment\_{duration} factor was missing until v4.5.0 despite FastSwitching being enabled since by default v4.0} so the mechanism only activates for the 8s and 15s latency targets as may be seen in Figure~\ref{fig:results-qoe-fs-on}.  Table~\ref{tab:exp-rerequests} shows that there are a significant number of re-requests which consume more bandwidth but don't generally increase the \ac{qoe} as may be seen in Figure~\ref{fig:results-qoe-fs-on}. We note that whilst the feature shows some benefit for the original L2A-LL algorithm at the 8s latency target (though with a much larger variance), there are some duplicate requests even at the 3s target as its buffer temporarily increases beyond the threshold. It also only manages to request 124.5 segments out of the 390 total, meaning it spent a significant portion of the time stalled.

We recommend that FastSwitching be avoided in low latency streaming.

\subsection{Analysis of earlier dash.js versions}

In the course of our studies we also analysed earlier versions of \texttt{dash.js}, so for example when using v4.3.0 the results seen in Figure~\ref{fig:results-qoe-d-430} (the 8s result was not available) shows clearer \ac{qoe} improvements for Dynamic. We observe that the performance results with \texttt{dash.js} v4.5.0 are not as differentiated as prior versions, though the ordering of the different algorithms is maintained. These results underscore the importance of carrying out performance testing of each version.

\section{Related Work}
\label{sec:related}

There is a large range of \ac{has} systems, a good overview of which may be found in survey papers~\cite{bentalebSurveyBitrateAdaptation2019, saniAdaptiveBitrateSelection2017}. 

There have been a number of studies of low latency \ac{has} systems, which have mainly examined the performance of individual low latency ABR algorithms at fixed low latencies, tending to focus on the lowest possible latency without consideration for performance at higher latencies \cite{karagkioulesOnlineLearningLowlatency2020,limWhenTheyGo2020}. 

One study \cite{zhangPerformanceLowLatencyDASH2022} analysed various low latency streaming players including the three \ac{abr} algorithms in the \texttt{dash.js} DASH player. Their tests were performed with a single fixed latency default (3s in the case of \texttt{dash.js}) target on the Mahimahi \cite{netravaliMahimahiAccurateRecordandReplay2015} network emulator using two 4G network profiles. They showed that Dynamic performed the best overall, although LoL and L2A-LL performed better at minimising the live latency, with L2A-LL outperforming LoL in average bitrate, but it experienced more rebuffering events. However, the average bitrates of their network profiles were an order of magnitude larger than their top test stream rendition bitrate which meant that the tests were less challenging and explored a more limited range of network conditions. 

There is a range of existing approaches for testing \ac{has} systems. The Mahimahi \cite{netravaliMahimahiAccurateRecordandReplay2015} system, which has been utilised by a number of studies \cite{maoNeuralAdaptiveVideo2017,zhangPerformanceLowLatencyDASH2022}, implements its own user-level network emulation environment which can be controlled by packet trace files. It utilises Linux’s tunnel interfaces and network namespaces to provide for session isolation. Sessions are hosted on virtual machines (VMs) with virtual network interfaces connecting them to the Mahimahi system. However, our approach utilises the Linux kernel traffic control and full network isolation by running each experiment on a separate physical machine.

Both Firefox and the Chrome based browsers contain network throttling functionality as part of their developer tools (DevTools) network module. The network throttling is performed at the application level so it doesn’t allow for network level emulation, which means it is less realistic than the approach we have taken.  The 2020 ACM Multimedia Systems Conference’s Grand Challenge \footnote{\url{https://2020.acmmmsys.org/lll_challenge.php}} provided a framework based on the Chrome browser’s network throttling functionality to provide for controlled network profiles. These network profiles consisted of limited lists of speed and duration, with a short number of throughput changes overall, which aren't such a good representation of the kind of fluctuations experienced on the public internet.

\section{Conclusions}
\label{sec:conclusions} 

In this paper we investigated how differing latency targets can affect the performance and \ac{qoe} in the \texttt{dash.js} player. We explored the performance across a range of realistic network profiles and different configuration settings. We found that with dash.js v4.5.0 the default Dynamic algorithm achieves the best overall QoE, though LoL and L2A-LL can achieve higher video quality at lower latencies, but suffer increased stalling. The Dynamic algorithm also manages to adhere to the target better than the other algorithms.

Further analysis of the poor performance of the L2A-LL algorithm enabled us to develop modifications to the implementation which showed a significant improvement.  We examined the FastSwitching algorithm and found that for low latency it provides no improvement in the overall QoE, and wastes bandwidth. Our investigations also discovered that there can be notable performance differences between software versions which highlights the need for testing specific versions before deployment and potentially before releases. 

This paper helps to inform the selection of the appropriate latency targets for low latency streaming with the current set of algorithms. It provides evidence that aiming for too low a latency target can impact QoE so providers need to find an appropriate compromise between latency and QoE. Furthermore, we elucidate examples that highlight the fact that providers need to perform sufficient analysis on the specific version and configuration of \texttt{dash.js} before deployment to ensure suitable QoE is attained. 

In future we would like to explore more approaches to improve the \ac{qoe} of low latency streaming, including analysing further configuration and algorithm improvements, and further exploring the L2A-LL improvements. We are also interested to investigate how a per-client optimum latency target can be determined which adapts to its network and own capabilities.

\bibliographystyle{ACM-Reference-Format}
\bibliography{ZoteroLibrary}

\end{document}